\documentstyle[prl,aps,tighten]{revtex}

\begin{document}

\twocolumn[
\hsize\textwidth\columnwidth\hsize\csname@twocolumnfalse\endcsname

\title{Reconstruction of a scalar-tensor theory of gravity\\
in an accelerating universe}

\author{B.~Boisseau$^a$, G.~Esposito-Far\`ese$^{b,c}$,
D.~Polarski$^{a,c}$, and A.A.~Starobinsky$^{d,e}$}

\address{$^a$ Lab.~de Math\'ematique et Physique Th\'eorique,
UPRES-A 6083 CNRS,\\
Universit\'e de Tours, Parc de Grandmont, F 37200 Tours, France}
\address{$^b$ Centre de Physique Th\'eorique, CNRS Luminy,
Case 907, F 13288 Marseille cedex 9, France}
\address{$^c$ D\'epartement d'Astrophysique Relativiste et de
Cosmologie,\\
Observatoire de Paris-Meudon, F 92195 Meudon cedex, France}
\address{$^{d}$ Landau Institute for Theoretical Physics,
117334 Moscow, Russia}
\address{$^{e}$ Newton Institute for Mathematical Sciences,
University of Cambridge, Cambridge CB3 0EH, U.K.}

\date{January 21, 2000}
\maketitle

\begin{abstract}
The present acceleration of the Universe strongly indicated by recent
observational data can be modeled in the scope of a scalar-tensor
theory of gravity. We show that it is possible to determine the
structure of this theory (the scalar field potential and the
functional form of the scalar-gravity coupling) along with the
present density of dustlike matter from the following two observable
cosmological functions: the luminosity distance and the linear
density perturbation in the dustlike matter component as functions of
redshift. Explicit results are presented in the first order in the
small inverse Brans-Dicke parameter $\omega^{-1}$.
\end{abstract}

\preprint{CPT-99/P.3917}

\draft

]

Recent observational data on type Ia supernovae explosions at high
redshifts $z\equiv {a(t_0)\over a(t)}-1\sim 1$ obtained independently
by two groups~\cite{Perl,Garn}, as well as numerous previous
arguments (see the recent reviews~\cite{BOPS,SS}), strongly support
the existence of a new kind of matter in the Universe whose energy
density not only is positive but also dominates the energy densities
of all previously known forms of matter [here $a(t)$ is the scale
factor of the Friedmann-Robertson-Walker (FRW) isotropic cosmological
model, and $t_0$ is the present time]. This form of matter has a
strongly negative pressure and remains unclustered at all scales where
gravitational clustering of baryons and (non-baryonic) cold dark
matter (CDM) is seen. Its gravity results in the present acceleration
of the expansion of the Universe: $\ddot a(t_0) >0$. In a first
approximation, this kind of matter may be described by a constant
$\Lambda$-term in the gravity equations as first introduced by
Einstein. However, a $\Lambda$-term could also be slowly varying with
time. If so, this will be soon determined from observational data. In
particular, if we use the simplest model of a variable $\Lambda$-term
(also called quintessence in~\cite{CDS}) borrowed from the
inflationary scenario of the early Universe, namely an effective
scalar field $\Phi$ with some self-interaction potential $U(\Phi)$
minimally coupled to gravity, then the functional form of $U(\Phi)$
can be determined from observational {\it cosmological functions}:
either from the luminosity distance $D_L(z)$~\cite{St98,HT}, or from
the linear density perturbation in the dustlike component of matter
in the Universe $\delta_m (z)$
for a fixed comoving smoothing radius~\cite{St98}.
However, this model cannot account for
{\it any} future observational data, in particular, for any
functional form of $D_L(z)$. This happens because a variable
$\Lambda$-term in this model should satisfy the weak-energy condition
$\varepsilon_{\Lambda}+p_{\Lambda}\ge 0$. In terms of the observable
quantity $H(z)\equiv {\dot a(t)/ a(t)}$ describing the evolution of
the expanding Universe at recent epochs,
the following inequality should be satisfied~\cite{SS}
\begin{equation}
\frac{dH^2(z)}{dz}\ge 3 \Omega_{m,0} H_0^2 (1+z)^2~.\label{ineq}
\end{equation}
Here, $H_0=H(z=0)$ is the Hubble constant,
$\Omega_{m,0}$ is the present energy density of the dustlike
(CDM+baryons) matter component in terms of the critical density
$\varepsilon_{\rm crit}=3H_0^2/8\pi G$ ($c=\hbar = 1$, and an index
$0$ stands for the present value of the corresponding quantity).
Note that the inequality~(\ref{ineq}) saturates when the
$\Lambda$-term is exactly constant. It is not clear from the
existing data whether~(\ref{ineq}) is satisfied at all. Actually the
opposite holds: An attempt to reconstruct $U(\Phi)$ from the
supernovae data~\cite{SSSS} and fitting of existing data to a model
with a linear equation of state for the $\Lambda$-term $p_{\Lambda}=
w\varepsilon_{\Lambda}$, with $w< -1$~\cite{Cald}, shows that the
possibility of violation of inequality~(\ref{ineq}), though strongly
restricted, is not completely excluded.
Hence it is natural and important to consider a variable
$\Lambda$-term in a more general class of scalar-tensor theories of
gravity where the requirement (\ref{ineq}) does not arise.
Moreover, this generalization of general relativity (GR) is inspired
by present more fundamental quantum theories, like $M$-theory. In
these theories, the scalar field $\Phi$ is just the dilaton field,
hence we shall call it so below.

Thus, we are interested in a universe where gravity is described by a
scalar-tensor theory, and we consider the Lagrangian density in the
Jordan frame~\cite{bnw70}
\begin{equation}
L={1\over 2} \Bigl (F(\Phi)~R -
g^{\mu\nu}\partial_{\mu}\Phi\partial_{\nu}
\Phi \Bigr) - U(\Phi) + L_m(g_{\mu\nu})~,
\label{L}
\end{equation}
where $L_m$ describes dustlike matter and $F(\Phi)>0$. This
corresponds to the Brans-Dicke parameter $\omega=F/(dF/d\Phi)^2
> 0$. One may also introduce a function $Z(\Phi)$ in front of the
kinetic term $(\partial_{\mu}\Phi)^2$, but it can be set either to
$1$, or to $-1$ by a redefinition of the scalar field. Under the
assumption of absence of ghosts in the theory, the second possibility
requires the Brans-Dicke parameter to lie in the range $-3/2 < \omega
<0$ (see~\cite{bep99} for more details). Since this clearly
contradicts solar system tests of GR either in the absence of
$U(\Phi)$, or for $U(\Phi)$ satisfying the condition (\ref{R}) below
for scales of galaxies and clusters of galaxies, we will not discuss
this possibility further. We do not introduce any direct coupling
between $\Phi$ and $L_m$ (though this possibility could be envisaged,
too). This guarantees that the weak equivalence principle is exactly
satisfied (universality of free-fall of laboratory-size objects), and
also that fundamental constants, like e.g. the fine-structure
constant, do not change with time in this theory. This is in very
good agreement with laboratory, geophysical and cosmological
data~\cite{s94,dd96,ipv}.

Such a scalar-tensor theory was recently
considered as a model for a variable $\Lambda$-term for some special
choices of $F(\Phi)$ and $U(\Phi)$ (see~\cite{PBM}). Our approach is
just the opposite: We want to {\it derive\/} these functions from
observational data.
Since we have to determine {\it two\/} functions $F(\Phi)$ and
$U(\Phi)$, we will need both observational functions $D_L(z)$ and
$\delta_m(z)$, in contrast to GR. Then the reconstruction problem
can be uniquely solved as will be shown below. Note that the angular
diameter as a function of $z$ provides the same information as
$D_L(z)$ (see~\cite{SS} and the second reference in~\cite{St98}).

It is most appropriate for us to work in the Jordan frame (JF), in
which the various physical quantities are those that are being
measured in experiments, even though the Einstein frame (EF) often
provides a better mathematical insight. In addition, the dilaton
appears to be directly coupled to dustlike matter in the EF frame, in
contrast to the JF. For a flat FRW universe with $ds^2= -dt^2 + a^2
d{\bf x}^2$, the background equations in the JF are then
\begin{eqnarray}
3FH^2 &=& \rho_m + {\dot \Phi^2\over 2} + U - 3H {\dot F}\ ,
\label{H2}\\
-2 F {\dot H} &=& \rho_m + \dot \Phi^2 + {\ddot F} - H {\dot F}\ .
\label{Hdot}
\end{eqnarray}
Their consequence is the equation for the dilaton itself:
\begin{equation}
\ddot \Phi +3H\dot\Phi +{dU\over d\Phi}-
3(\dot H + 2H^2)~{dF\over d\Phi} =0~.
\end{equation}
Combining Eqs.~(\ref{H2})--(\ref{Hdot}) and changing the
argument from time $t$ to redshift $z$, we obtain the
following basic equation for $F(z)$:
\begin{eqnarray}
F'' &+& \left[(\ln H)' - \frac{4}{1+z}\right]~F' +
\left[\frac{6}{(1+z)^2} - \frac{2(\ln H)'}{1+z}\right]~F
\nonumber\\
&=& \frac{2U}{(1+z)^2 H^2} + 3~(1+z)
\left(\frac{H_0}{H}\right)^2 F_0~\Omega_{m,0}\ ,
\label{F}
\end{eqnarray}
where the prime denotes the derivative with respect to $z$.

The effective value of Newton's gravitational constant $G_N$ in
Eqs.~(\ref{H2}--\ref{Hdot}) is given by the formula $G_N=1/8\pi F$.
We shall use its present value $G_{N,0}$ in the definition of the
critical density
$\varepsilon_{\rm crit}$. On the other hand, $G_{N,0}$ is {\it not\/}
the quantity measured in laboratory Cavendish-type and solar-system
experiments. For a massless dilaton, the effective gravitational
constant between two test masses is given by
\begin{equation}
G_{\rm eff} = {1\over 8\pi F}
\left({2F+4(dF/d\Phi)^2\over 2F+3(dF/d\Phi)^2}\right)~.
\label{Geff}
\end{equation}
In our case, the dilaton is massive, so the expression (\ref{Geff})
will be valid for physical scales $R$ such that
\begin{equation}
R^{-2} \gg \max \left(\left|{d^2U\over d\Phi^2}\right|, H^2,
H^2 \left|{d^2F \over d\Phi^2}\right|\right).
\label{R}
\end{equation}
Previously, the expression $G_{\rm eff}$ was known from the
post-Newtonian expansion; below we rederive it using the
cosmological perturbation theory.

Let us now list the restrictions of the theory (\ref{L}) which
follow from solar-system and cosmological tests. The post-Newtonian
parameters $\beta$ and $\gamma$ for this theory are:
\begin{eqnarray}
\gamma &=& 1 - {(dF/d\Phi)^2\over F + 2(dF/d\Phi)^2}~,
\label{gammaPPN}\\
\beta &=& 1 + {1\over 4}\,{F~(dF/d\Phi)\over 2 F + 3(dF/d\Phi)^2}\,
{d\gamma\over d\Phi}\ .
\label{betaPPN}
\end{eqnarray}
Using the upper bounds on $(\gamma-1)$ from solar system
measurements~\cite{vlbi,dw96}, we get
\begin{equation}
\omega_0^{-1}=F_0^{-1}~(dF/d\Phi)_0^2 < 4\times 10^{-4}~.
\label{dF0}
\end{equation}
So, $G_{N,0}$ and $G_{{\rm eff},0}$ coincide with better than
$2\times 10^{-4}$ accuracy. On the other hand, the difference between
$G_N$ and $G_{\rm eff}$ may be larger at redshifts $z\sim 1$ since
neither the upper limit on $\beta$, nor the present experimental
bound $|\dot G_{\rm eff}/G_{\rm eff}|< 6\times 10^{-12}\ {\rm
yr}^{-1}$~\cite{dw96} significantly restrict $(d^2F/d\Phi^2)_0$.
Note that we cannot use the nucleosynthesis bound on the change of
$G_{\rm eff}$ since that time as the behavior of $G_{\rm eff}$ during
the intermediate period is unknown, unless we make additional
assumptions (see below).

The theory (\ref{L}) describes a variable $\Lambda$-term with desired
properties if the following three conditions are satisfied:

1) The $\Lambda$-term is dynamically important at present, namely,
$\Omega_{\Lambda,0}\sim 0.7 \sim 2 \Omega_{m,0}$, or
\begin{equation}
\left({\dot\Phi^2\over 2}+U-3H\dot F\right)_0\sim 0.7
\varepsilon_{\rm crit}\sim 2\rho_{m,0}~.
\label{1st}
\end{equation}

2) The $\Lambda$-term has a sufficiently large negative pressure to
provide acceleration of the present Universe. The condition
$\ddot a_0 > 0$ reads:
\begin{equation}
2U_0>(\rho_m+2\dot\Phi^2+3\ddot F+3H\dot F)_0~.
\label{2st}
\end{equation}

3) The dark matter described by the $\Lambda$-term remains
unclustered at scales up to $R\sim 10h^{-1}(1+z)^{-1}$ Mpc and
probably even more (here $h=H_0/100$ km s$^{-1}$ Mpc$^{-1}$). To
achieve this, it is {\it sufficient\/} to assume that the
inequality~(\ref{R}) is satisfied for all scales in question.

The first step of our program is purely kinematical:
we determine $H(z)$ from $D_L(z)$ like in GR,
\begin{equation}
{1\over H(z)} = \left(\frac{D_L(z)}{1+z} \right)'~.
\end{equation}
The functional dependence of $D_L(z)$ on the cosmological parameters,
like $\Omega_{m,0}$, is of course model dependent. If $\Omega_{m,0}$
is already known from other tests, we can find already at that stage
of the reconstruction a quantity such as the present effective
equation of state of the dilaton from the formula (cf.~\cite{SSSS}):
\begin{equation}
w_0\equiv \frac{p_{\Lambda,0}}{\varepsilon_{\Lambda,0}}=
{(2/3)(d\ln H/dz)_0-1\over 1- \Omega_{m,0}}~.
\label{w0}
\end{equation}
$\varepsilon_{\Lambda,0}$ contains the term $-3H_0\dot F_0$, so that
$\Omega_{m,0}+\Omega_{\Lambda,0}=1$. The dilaton equation of state
can be determined for $z>0$, too; one has only to define what should
be called the pressure and the energy density of the dilaton in
general. Actually, we will show below that
$\Omega_{m,0}$ is itself self-consistently determined from our
approach, so no additional information is required to find $w(z)$.

In contrast to GR, Eq.~(\ref{F}) is no longer sufficient to
determine $U(z)$; one should know $F(z)$, too. For this purpose we
will use $\delta_m(z)$. We consider perturbations in the longitudinal
gauge
$ds^2= -(1 + 2 \phi) dt^2 + a^2 (1 - 2\psi) d{\bf x}^2$.
Working in Fourier space (a spatial dependence
$\exp(i{\bf k}.{\bf x})$ with $k\equiv |{\bf k}|$ is assumed), the
following equations are obtained:
\begin{eqnarray}
\phi &=& \dot v = \psi - \delta F/F\ ,\label{cons1}\\
\dot\delta_m &=&
-\frac{k^2}{a^2}\, v + 3 {d(\psi + H v)\over dt}~,\label{cons2}
\end{eqnarray}
where the gauge invariant quantity
$\delta_m \equiv (\delta\rho_m)/\rho_m + 3Hv$, and $v$ is the
peculiar velocity potential of dustlike matter. We also get
\begin{eqnarray}
-3 \dot F \dot\phi &-& \left( 2{k^2\over a^2} F - \dot\Phi^2 +
3H\dot F \right)\phi =
\nonumber\\
&=& \rho_m\delta_m + 3{\dot F\over F}\,{\dot{\delta F} }
+ \left( {k^2\over a^2}-6H^2 -3{\dot F^2\over F^2} \right)\delta F
\nonumber\\
&+& \dot\Phi \dot{\delta\Phi}
+ 3H\dot\Phi\,\delta\Phi
+ \delta U , \label{flmetr}
\end{eqnarray}
and the equation for the dilaton fluctuations $\delta\Phi$:
\begin{eqnarray}
\ddot{\delta\Phi} + 3H\dot{\delta\Phi}
&+& \left[{k^2\over a^2} - 3(\dot H + 2H^2){d^2F\over d\Phi^2}
+{d^2U\over d\Phi^2}\right]\delta\Phi =
\nonumber\\
&=& \left[{k^2\over
a^2}(\phi-2\psi)-3(\ddot\psi+4H\dot\psi+H\dot\phi)\right]{dF\over
d\Phi}
\nonumber\\
&& + (3\dot\psi+\dot\phi)\dot\Phi - 2\phi{dU\over d\Phi}\ .
\label{fluctPhi}
\end{eqnarray}

Let us now consider sufficiently small scales $R=2\pi a/k$ for which
the inequality (\ref{R}) is well satisfied. For
example, if
$\delta_m(z)$ is determined from the abundance of rich clusters of
galaxies, then the relevant comoving scale is $R\sim
8h^{-1}/(1+z)$~Mpc. If the r.h.s. of Eq.~(\ref{R}) is $\sim
H^2$, then the corresponding small parameter is $R^2H_0^2\sim
10^{-5}$. Note that we have another parameter, $\omega^{-1}$,
which is small at the present time, Eq.~(\ref{dF0}),
but it need not be so small in the past. Also, this parameter may be
larger than $a^2H^2/k^2$. For this reason, we will first keep it.

The solution of Eqs.~(\ref{cons1}--\ref{fluctPhi}) in the formal
short-wavelength limit $k\to \infty$ can be found following the
analytical method used in~\cite{St98} in the GR case, confirmed
numerically in~\cite{Ma}. The idea is that the leading terms in
Eqs.~(\ref{cons1}--\ref{fluctPhi}) are either those containing $k^2$,
or those with $\delta_m$.
Then, using (\ref{cons2}) and
the l.h.s. of Eq.~(\ref{cons1}), the standard form of the equation for
dustlike matter density perturbation follows:
\begin{equation}
\ddot \delta_m + 2H\dot \delta_m + k^2a^{-2}\phi \simeq 0~.
\label{B}
\end{equation}
Now we consider the solution of Eq.~(\ref{fluctPhi}) of interest
to us, for which $|\ddot{\delta \Phi}|\ll k^2a^{-2}|\delta\Phi|$.
It corresponds to the growing adiabatic mode. So, keeping terms
with $k^2$ in Eq.~(\ref{fluctPhi}) and then using the r.h.s. of
Eq.~(\ref{cons1}), we obtain:
\begin{equation}
\delta \Phi \simeq (\phi - 2\psi)~{dF\over d\Phi} \simeq
- \phi~{F~dF/d\Phi\over F+2(dF/d\Phi)^2}~.
\label{dPhi}
\end{equation}
In the GR case, $\delta \Phi \propto k^{-2}\phi$ in the limit $k\to
\infty$, so matter producing the $\Lambda$-term is not gravitationally
clustered at small scales (physically, due to free streaming). This is
not so in scalar-tensor gravity: The dilaton remains partly clustered
for arbitrarily small scales, this clustering being small only because
$\omega$ is large.

Keeping only terms with $k^2$ or $\delta_m$ in Eq.~(\ref{flmetr}), we
get the expression of $\phi$ through $\delta_m$ and $\delta F$.
Finally, inserting it into Eq.~(\ref{B}) and using Eq.~(\ref{dPhi}),
we arrive to the closed form of the equation for $\delta_m$:
\begin{equation}
{\ddot \delta_m} + 2H {\dot \delta_m} - 4\pi G_{\rm eff}\,
\rho_m~\delta_m\simeq 0~,\label{del}
\end{equation}
with $G_{\rm eff}$ defined in (\ref{Geff}) above. In terms
of $z$, (\ref{del}) reads:
\begin{eqnarray}
H^2~\delta_m'' + \left(\frac{(H^2)'}{2} -
{H^2\over 1+z}\right)\delta_m'
\nonumber \\
\simeq {3\over 2} (1+z) H_0^2 {G_{\rm eff}(z)\over
G_{N,0}}~\Omega_{m,0}~\delta_m~.\label{delz}
\end{eqnarray}
Eq.~(\ref{del}) does not contain $k^2$ at all. Thus, its solutions,
as well as the corresponding expressions for $\delta \Phi$, do not
oscillate with the frequency $k/a$ for $k\to \infty$. This justifies
the assumption about $\ddot {\delta \Phi}$ made above.

Extracting $H(z)$ (from $D_L(z)$) and $\delta_m(z)$ from observations
with sufficient accuracy, we can reconstruct $G_{\rm eff}(z)/G_{N,0}$
analytically. Since, as follows from Eq.~(\ref{dF0}), the quantities
$G_{{\rm eff},0}$ and $G_{N,0}$ coincide with better than $0.02\%$
accuracy, Eq.~(\ref{delz}) taken at $z=0$ gives also the value of
$\Omega_{m,0}$ with the same accuracy. Thus, in principle, no
independent measurement of $\Omega_{m,0}$ is required.

The resulting equation $G_{\rm eff}(z)=p(z)$, where $p(z)$ is a given
function following from observational data, can be transformed into a
nonlinear second order differential equation for $F(z)$ if we exclude
$d\Phi$ (which appears in $dF/d\Phi$) using the background equation
(\ref{Hdot}), which reads
\begin{eqnarray}
\Phi'^2 &=& - F'' - \left[(\ln H)' + \frac{2}{1+z}\right]~F'
\nonumber\\
&& +\frac{2(\ln H)'}{1+z}~F
-3(1+z) \frac{H_0^2}{H^2} F_0~\Omega_{m,0}~.
\label{Phi}
\end{eqnarray}
Therefore, $F(z)$ can be determined by solving that equation provided
$F_0$ ($=1/8\pi G_{N,0})$ and $F'_0$ are known.

However, this procedure can be simplified a lot under reasonable
assumptions, and taking into account the small present values of
$\omega^{-1} = F^{-1}(dF/d\Phi)^2$ and $\dot G_{\rm eff}/G_{\rm
eff}$. Indeed, the value of $\omega^{-1}$ for $0\le z \alt 1$ can
be estimated from the first terms of its Taylor expansion
$\omega^{-1}_0 + z\, (d\omega^{-1}/dz)_0$. Neglecting contributions
proportional to $\omega^{-1}_0$, we then get
$\omega^{-1} \sim 2\, z\, \lambda\, (d^2F/d\Phi^2)_0$,
with $\lambda \equiv -(d\ln F/d\Phi)_0\, \dot\Phi_0/H_0$,
whereas $\dot G_{\rm eff}/G_{\rm eff} \simeq \lambda\, H_0\,
[1-(d^2F/ d\Phi^2)_0]$. If $(d^2F/d\Phi^2)_0$ differs
significantly from 1, we can thus conclude that
$\omega^{-1} \alt |2\dot G_{\rm eff}/H_0 G_{\rm eff}| \alt 0.25$.
On the other hand, if $(d^2F/d\Phi^2)_0$ happens to be close to 1,
one can still assume that there is no special cancellation of large
terms in the r.h.s. of Eq.~(\ref{H2}), and therefore that
$\dot\Phi_0^2\alt 6 F_0 H_0^2$. The above estimate for $\omega^{-1}$
then gives $\omega^{-1} \alt 2\sqrt{6/\omega_0} \alt 0.1$. In both
cases, we thus find that
$G_{\rm eff}\simeq G_N$ in the range of $z$ involved with better
than $\sim 10\%$ accuracy. Note that the same estimate may be
obtained by assuming that $\omega^{-1}$ changed {\it monotonically\/}
with $z$ and using the nucleosynthesis bound (cf.~\cite{PBM}).
Therefore, in first approximation in $\omega^{-1}$,
$G_{\rm eff}(z) \simeq 1/8\pi F(z)$ and Eq.~(\ref{delz}) can be used
to determine $F(z)$ unambiguously. Small corrections to this result
can be taken into account using perturbation theory with respect to
the small parameter $\omega^{-1}$. After $F(z)$ is found, the
potential $U(z)$ is determined from Eq.~(\ref{F}).

Finally, using Eq.~(\ref{Phi}) we find $\Phi(z)$ by simple
integration. After that, both unknown functions $F(\Phi)$ and
$U(\Phi)$ are completely fixed as functions of $\Phi-\Phi_0$ in that
range probed by the data. Equations (\ref{delz}), (\ref{F}) and
(\ref{Phi}), giving the subsequent steps of the reconstruction,
constitute the fundamental result of our letter.

Our results generalize those obtained in GR~\cite{St98} and
constrain any attempt to explain a varying
$\Lambda$-term using scalar-tensor theories of gravity.
Good data on $\delta_m(z)$ expected to appear soon from observations
of clustering and abundance of different objects at redshifts $\sim
1$ and more, as well as from weak gravitational lensing, together
with better data on $D_L(z)$ from more supernova events, will allow
implementation of the reconstruction program and determination of
the microscopic Lagrangian.

\acknowledgments
A.S. was partially supported by the Russian Foundation for Basic
Research, grant 99-02-16224, and by the Russian Research Project
``Cosmomicrophysics''. Centre de Physique Th\'eorique is Unit\'e
Propre de Recherche 7061.

\end{document}